# Magnetocaloric effect near room temperature in chromium telluride (Cr$_2$Te$_3$)

*Nishant Tiwari[1] ‡, Chinmayee Chowde Gowda[2] ‡, Subhendu Mishra[3], Prafull Pandey[4], Saikat Talapatra[5], Abhishek K. Singh[3]\* and Chandra Sekhar Tiwary[1,2]\**

[1]Department of Metallurgy and Materials Engineering, Indian Institute of Technology Kharagpur, West Bengal - 721302, India

[2]School of Nano Science and Technology, Indian Institute of Technology Kharagpur, West Bengal – 721302, India

[3]Materials Research Centre, Indian Institute of Science, Bengaluru, Karnataka – 560012, India

[4]Department of Materials Engineering, Indian Institute of Technology Gandhinagar, Gujarat – 382355, India

[5]School of Physics and Applied Physics, Southern Illinois University, Carbondale, IL 62901, USA

ABSTRACT Transition metal telluride compositions are explored extensively for their unique magnetic behavior. Since chromium telluride ($Cr_2Te_3$) exhibits a near room temperature phase transition, the material can be effectively used in applications such as magnetic refrigeration. Compared to existing magnetocaloric materials, Heusler alloys and rare-earth based alloys, the large-scale synthesis of $Cr_2Te_3$ involves less complexity, resulting in a stable composition. Compared to existing tellurides, $Cr_2Te_3$ exhibited a large magnetic entropy change ($\Delta S_M$) of 2.36 J/kg-K at a very small magnetic field of 0.1 T. The refrigeration capacity (RC) of ~ 160 J/kg was determined from entropy change versus temperature curve. The results were comparable with the existing Cr-compounds. The telluride system, $Cr_2Te_3$ compared to pure gadolinium, reveals an enhanced room temperature magnetocaloric effect (MCE) with broad working temperature range. The heating cycle of MCE was successfully visualized using a thermal imaging setup. To confirm the observed magnetic properties of $Cr_2Te_3$, first-principles calculations were conducted. Through density functional theory (DFT) studies, we were able to determine both Curie temperature ($T_C$) and Néel temperature ($T_N$) which validated our experimental transitions at the same temperatures. Structural transition was also observed using first principles DFT calculation which is responsible for magnetic behavior.

## 1. Introduction

On a more recent note, 2D materials, including transition metal chalcogenides (TMDCs), received interest due to their excellent magnetic properties with potential applications in magnetocaloric, spintronics, topological, and anomalous Hall Effect. Telluride compositions varying from bulk to lower dimensions have gained wide attention due to their structural effects on magnetic property variation and their potential application in various fields[1–3]. The van der Waals ferromagnetic properties exhibited especially by the Cr-based chalcogenides sparked interest in reduced dimensional magnetic systems[4–10]. Few binary Cr-Te compositions studied in this regard are $Cr_4Te_5$ [4], $Cr_5Te_8$[5–7] and $Cr_2Te_3$[11]. These compounds are reported to be ferromagnets with the various magnetic phase transition temperature ranging from 160 to 340 K depending on Cr concentration. Addition of transition metal alloys in a magnetic matrix have been studied extensively for tuning the transition temperature to enable them for magnetocaloric and other thermal management application[12]. Recently, there have been ample studies on $Cr_2Te_3$, where the studies indicated that the magnetic phase transition temperature can also be tuned close to room temperature depending on the processing condition[13–16]. Due to its antiferromagnetic properties, studies have shown the addition of Cr can be used to tune the Curie temperature ($T_C$) near room temperature. Only a mere 5 % addition of Cr in Fe-Ni-Cr compositions resulted in the reduction of $T_C$ from 438 K to 258 K enabling uses near room temperature thermal management systems. Cr nanoparticles distributed in the ferrofluid improves thermal contact and accelerates heat exchange compared to bulk systems [17].

One such area of application for near room temperature (RT) magnetic phase transition is in the fields on non-gaseous magnetic refrigeration. Magnetic refrigeration which works on magnetocaloric effect (MCE) of magnetic materials has reconstituted refrigeration industry due to its energy efficient and a sustainable approach[18,19]. There are different types of magnetocaloric materials (MCM) which can be used for magnetic refrigeration which exhibits drop in temperature on application of external magnetic fields. MCE was first reported in gadolinium (Gd)[20], there after numerous rare earth-based alloys have been explored so far for their magnetocaloric properties[20,21]. Apart from rare earth-based alloys, Heusler alloys are also known for their giant magnetocaloric effect (GMCE)[22–24]. Many TMDCs with first order ferromagnetic to paramagnetic transitions exhibit MCE. Few telluride composition such as $Nd_3Te_4$ has also been explored for its enhanced magnetocaloric properties[25]. MCE in magnetocaloric materials (MCM) is quantified by the isothermal entropy change ($\Delta S_M$) or adiabatic temperature change ($\Delta T_{ad}$) with respect to applied magnetic field. Both parameters ($\Delta S_M$, $\Delta T_{ad}$) shows that maximum value around the magnetic transition temperature. Development of advanced MCMs have become an important research area since it plays a significant role in improving efficiency of magnetic refrigeration and an economic impact on the commercialization[26,27].

In the present study, we explore the magnetocaloric properties of $Cr_2Te_3$ synthesized using easily scalable melting process. As $Cr_2Te_3$ exhibits a second order magnetic phase transition which provides the entropy change due to magnetic field over broad temperature ranges and with no magnetic or thermal hysteresis, the material is best suited for magnetocaloric application. The magnetocaloric properties of $Cr_2Te_3$ sample was thoroughly analyzed using magnetic measurements as well as thermal analysis. Temperature dependent magnetization (M-T) measurements show the paramagnetic to antiferromagnetic transition ($T_N$) in the range of 288 - 294 K. To further validate these experiments, we have also fabricated a magnetocaloric material device and analyzed the thermal fluctuations with and without the presence of magnetic fields with the help of a thermal imaging camera. To further verify our experimental findings, density functional theory (DFT) calculations were also performed to explore the relationship between experimental and calculated $T_C$ and $T_N$ data of $Cr_2Te_3$. DFT calculations were performed to simulate structural changes in the investigated alloys at different temperatures using *ab-initio* Molecular Dynamics (AIMD).

## 2. Material and methods

### 2.1 Material synthesis

$Cr_2Te_3$ alloy was prepared by using induction melting, due to large temperature difference between Cr and Te, stochiometric amounts of the elements are sealed and melted [28]. Further the heating temperature was kept below the boiling point of Te (1223 K) and increased holding time (4 hours) to ensure time dependent diffusion. In order to get the uniform heating, sealed sample heated from the both (up and down) sides at same temperature as well as time. The prepared sample of $Cr_2Te_3$ as-cast was homogenized at 1173 K for 100 hours followed by furnace cooling to make sure heating and cooling rate must be same.

### 2.2 Characterization tools

The compositions of the prepared sample were confirmed using energy dispersive X-ray spectroscopy (EDS). The X-ray diffraction pattern is collected at room temperature using Cu-K$_\alpha$ target in Bruker D8 advance diffractometer. The microstructure features were analyzed using ZEISS GEMINI 600 field-emission scanning electron microscope (FE-SEM). The structural characterization was further probed using a high-angle annular dark-field scanning transmission electron microscopy (HAADF-STEM) operated at 300 kV. The magnetic characterization of the samples is performed using the SQUID (Superconducting quantum interference devices) module of a MPMS (Quantum Design, USA). The magnetization versus temperature (M-T) behavior of the samples were analyzed in conditions such as Field cooled warming (FCW) and Field cooled cooling (FCC) protocols at 0.1 T (Tesla) magnetic field. The differential scanning calorimetry (DSC) measurements have been carried out to validate the transition temperature within expected range of different transitions in both heating and cooling modes in Nitrogen atmosphere. Furthermore, M-H behavior of investigated alloys is measured across transition temperature ($T_N$) by varying the field from 0 - 4 T to calculate the entropy change ($\Delta S_M$) due to varying magnetic field.

**2.3 Device fabrication and Thermal imaging**

A device was fabricated with $Cr_2Te_3$ powder mixed with acetone and then molded into a thin layer of thickness 0.5 mm. The device was heated at 60 °C to eliminate acetone and moisture from the material. Later copper tape was attached at the end to verify the same effects on a commonly used conducting material, which usually serves as a contact for electrical device units. The refrigerant material device was exposed to low magnetic fields of 1.32 G. We used a thermal imaging setup to observe the heat fluctuations during heating and cooling cycle. The thermal imaging camera (Optris PI 640i) with temperature range from – 35 °C (238 K) till 40 °C (313 K).

## 2.4 Computational Methodology

The Vienna *ab initio* simulation package (VASP) [29,30] is used to perform the first-principles density functional theory (DFT) calculations. The projector-augmented wave (PAW) potentials are used to describe the ion-electron [31,32] interactions in the systems. The electronic exchange and correlation part of the potential is represented by the Perdew-Burke-Ernzerhof (PBE) [33] generalized gradient approximation (GGA). The Kohn Sham orbitals are expanded using plane wave basis sets with a 500 eV energy cutoff. The conjugate-gradient algorithm is used to relax all structures until the Hellmann-Feynman forces on each atom are less than 0.005 eVÅ$^{-1}$. For sampling the Brillouin zone (BZ) of $Cr_2Te_3$, a well-converged Γ−centred Monkhorst-Pack (MP) [34] k-grids of $12 \times 12 \times 1$. We perform *ab initio* Molecular Dynamics (AIMD) analysis at different constant temperature values (using the Langevin thermostat within the canonical ensemble (NPT) [35,36] to estimate the temperatures ($T_M$, $T_C$) where magnetic and structural phase transitions occur, respectively. The AIMD simulations are run for 2 ps with a time step of 1 fs. For AIMD calculations, we considered the $2 \times 2 \times 1$ supercells for $Cr_2Te_3$.

## 3. Results and Discussion

### 3.1 Structural evaluation

The induction melted $Cr_2Te_3$ samples formed a single-phase compound, as confirmed by X-ray diffraction. **Figure 1(a)** revealed a hexagonal Ni-As structure[37] and P$\bar{3}$1c (163) space group with ordered Cr vacancies. Cr and Te layers alternate in the layered unit cell, and Cr vacancies can be found in every other metal layer. The diffraction peaks correspond to the hexagonal $Cr_2Te_3$ phase and major planes were indexed as (110), (01$\bar{3}$), (11$\bar{2}$), (001), (021), (030), (116), (222) and (224) with JCPDS card number 98-001-5039. **Figure S1a** shows the Rietveld refinement of the bulk $Cr_2Te_3$ sample along with its peaks list matched with the above mentioned JCPDS card (98-001-5039) in **Figure S1b**. **Figure 1(a) inset** shows the polyhedral hexagonal crystal structure of the $Cr_2Te_3$. The samples' morphology was further investigated using SEM and elemental distribution of $Cr_2Te_3$ after 100 hours of homogenization. **Figure 1(b) inset** shows the EDS composition distribution with Cr: 21.05 wt% and Te: 78.64 wt% present in the compound. HAADF-STEM image of $Cr_2Te_3$ was further examined to determine the atomic arrangement pattern of the Cr and Te atoms. The atomic arrangements revealed two distinct domain orientations, both of which were hexagonal crystal systems, as illustrated in **Figure 1(c)**. The bright dots correspond to Te atoms, which have a larger atomic radius than chromium with a lattice spacing of 0.206 nm, corresponding to the $(1\bar{1}0)$ plane mapped in **Figure 1(a)** during XRD evaluation. The FFT pattern extracted from the HAADF STEM image (**Figure 1(c)**) depicts hexagonal phase formation. We further explore the sample composition confirmation through XPS analysis. **Figure 1(d)** shows Cr composition with peaks of Cr2p at 574 eV and 584 eV, as well as satellite peaks seen at 577 eV ($2p_{3/2}$) and 586 eV ($2p_{1/2}$), with FWHMs of 3.1 and 3.5, respectively. **Figure 1(e)** shows XPS of Te composition displaying Te3d peaks seen at 587.73 eV for $3d_{1/2}$ and 577.35 for $3d_{5/2}$, which coincide with the $Cr2p_{3/2}$ peak.

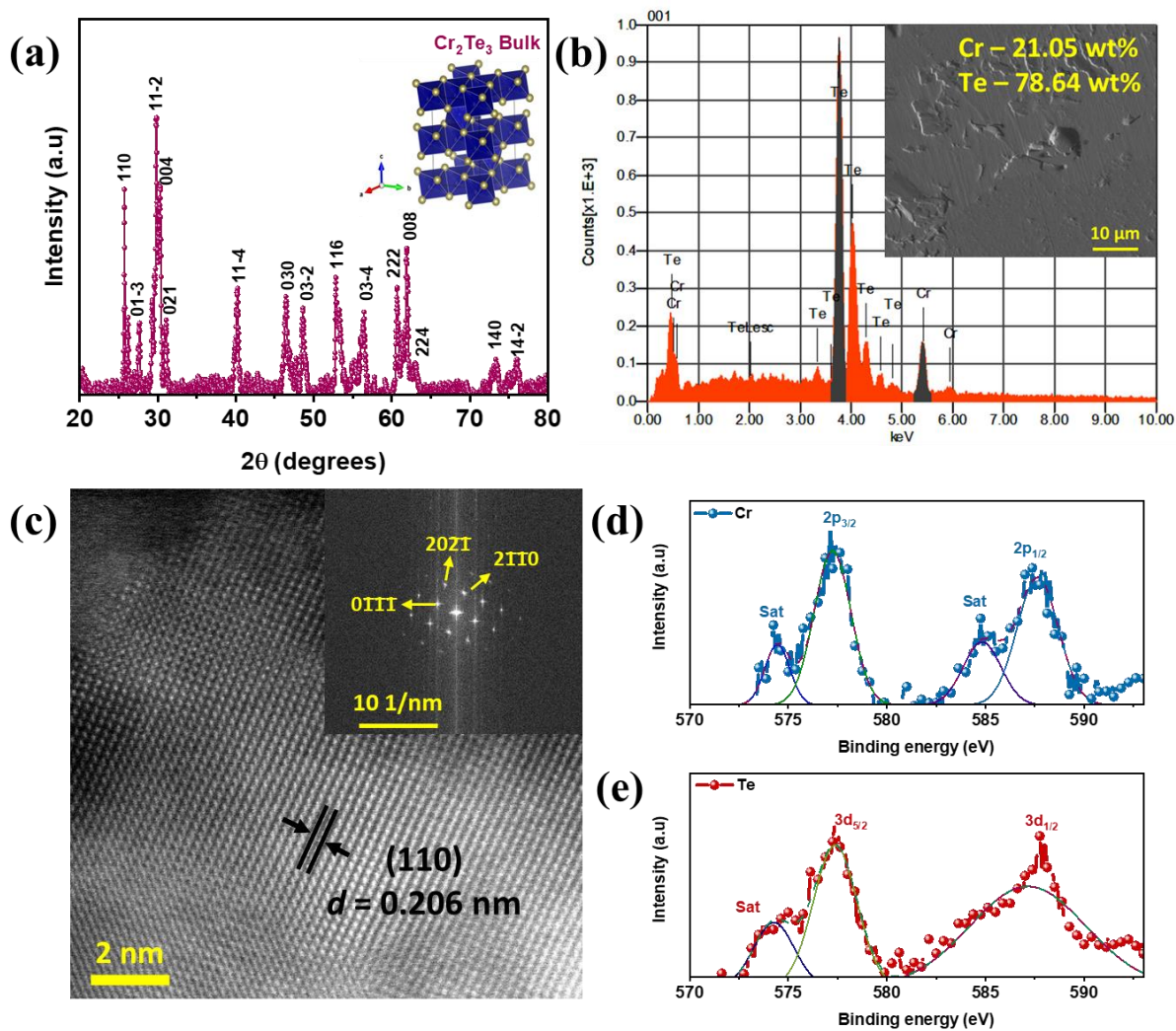

*Figure 1. (a) XRD profile of bulk $Cr_2Te_3$; inset: hexagonal crystal unit, (b) SEM microstructure along with the EDS (Inset) of $Cr_2Te_3$, (c) HAADF-STEM image of $Cr_2Te_3$ with d spacing of martensite ($L1_0$); inset: FFT pattern derived from (c), XPS spectra of $Cr_2Te_3$ sample (d) Cr and (e) Te composition.*

## 3.2 Magnetic property discussion

To determine the magnetic properties of $Cr_2Te_3$, measurements were conducted using a superconducting quantum interference device (SQUID). In order to explore the magnetic properties, the temperature dependent magnetization measurements (M-T) were performed with field warmed (FW) and field cooling (FC) condition under magnetic field of 0.1 T (Tesla) as shown in **Figure 2(a)**. The characteristic transformation temperature was determined to be 184.5 K where ferromagnetic to paramagnetic transition, Curie temperature ($T_C$) was observed. Further, near room temperature in the range of 288 - 292 K, paramagnetic to antiferromagnetic (PM to AFM) transition occurs which is of particular interest for magnetic refrigeration at room temperature. Along with this, first order differentiation of magnetization (dM/dT) is also plotted as a function of T (K) as shown in **Figure 2(b)**. A sharp drop in curve shows the $T_C$ around 184 K and a broad change in peak over a range nearby room temperature confirm the $T_N$ value as mentioned in **Figure 2 (b)**. This PM to AFM transformation temperature, Néel temperature ($T_N$) is consistent with DSC measurements as shown in **Figure 2(c)**. The second order transition occurring at a higher temperature around 290 K was paramagnetic to antiferromagnetic transition in bulk $Cr_2Te_3$. Paramagnetic to antiferromagnetic transitions have been observed in transition metal compounds at a narrow temperature range near Néel temperature ($T_N$) due to increase in temperature resulting in loss of magnetic ordering in the material. Few of the compounds include $FeBr_2$[38], manganites[39], $Cr_2O_3$[40] and other Cr compounds [17,41,42]. The transition in the near vicinity of critical temperature has been studied through specific-heat measurements. We confirm that the first order transition occurring at 184.5 K is ferromagnetic to paramagnetic transition accordingly. The experimental results were supported by DFT calculations showing first and second order magnetic transitions at similar temperatures in the further sections discussed. Following the similar way, first order differentiation of heat flow (dH/dT) has also been plotted to confirm the magnetic phase transition ($T_N$) as shown in heating and cooling cycle in **Figure 2(d)**. Further to confirm AFM behavior of $Cr_2Te_3$, isothermal magnetization (M-H) curve was measured at 300 K (T > $T_C$), which exhibits a linear behavior with small slope which is the characteristic nature of AFM as shown in **Figure 2(e)**. Half-metallicity in Cr compounds is caused by the ordering of the magnetic moment, which is localized at Cr sites and produces magnetism [43], $Cr_2Te_3$ is a metallic ferromagnet [44]. Four Cr sites are vertically oriented, and one-layer exhibits modest antiferromagnetic characteristics. Magnetic access with perpendicular magnetic field alignment is provided via the hexagonal c-axis. The average values of the Cr localized moment and Te's induced magnetic polarization are found to be antiparallel, at Cr = 3.30 µB/Cr and 0.18 µB/Te, respectively [44].

To determine the entropy change ($\Delta S_M$) due to magnetic field, across the ($T_N$), M-H isotherms (constant temperature) were measured from the 270 to 320 K within the range of 0-4 T magnetic field, as shown in **Figure 2(e)**. Magnetocaloric properties of all the investigated $Cr_2Te_3$ evaluated using integral of M-T curve and isothermal magnetization curve (M-H) as described in **Equation 1**.

$$\Delta S_M(T, H_{applied}) = \int_0^{H_{applied}} \left(\frac{\partial M}{\partial T}\right)_H dH \tag{1}$$

As shown in **Figure 2(e)**, $\Delta S_M$ is calculated using equation for the field change from 0 to 4T (Tesla). Owing to the magnetic phase transition from paramagnetic to antiferromagnetic the maximum $\Delta S_M$ is around 2.36 J/kg-K at 293 K. As one can see in **Figure 2(f),** The wide working temperature range directly influences the material's performance in magnetocaloric applications. This is known as cooling capacity or refrigeration capacity (RC), which is 160 J/Kg for the $Cr_2Te_3$ using **Equation 2.**

$$RC = \Delta S_M^{Max} \times (\Delta T_{FWHM}) \tag{2}$$

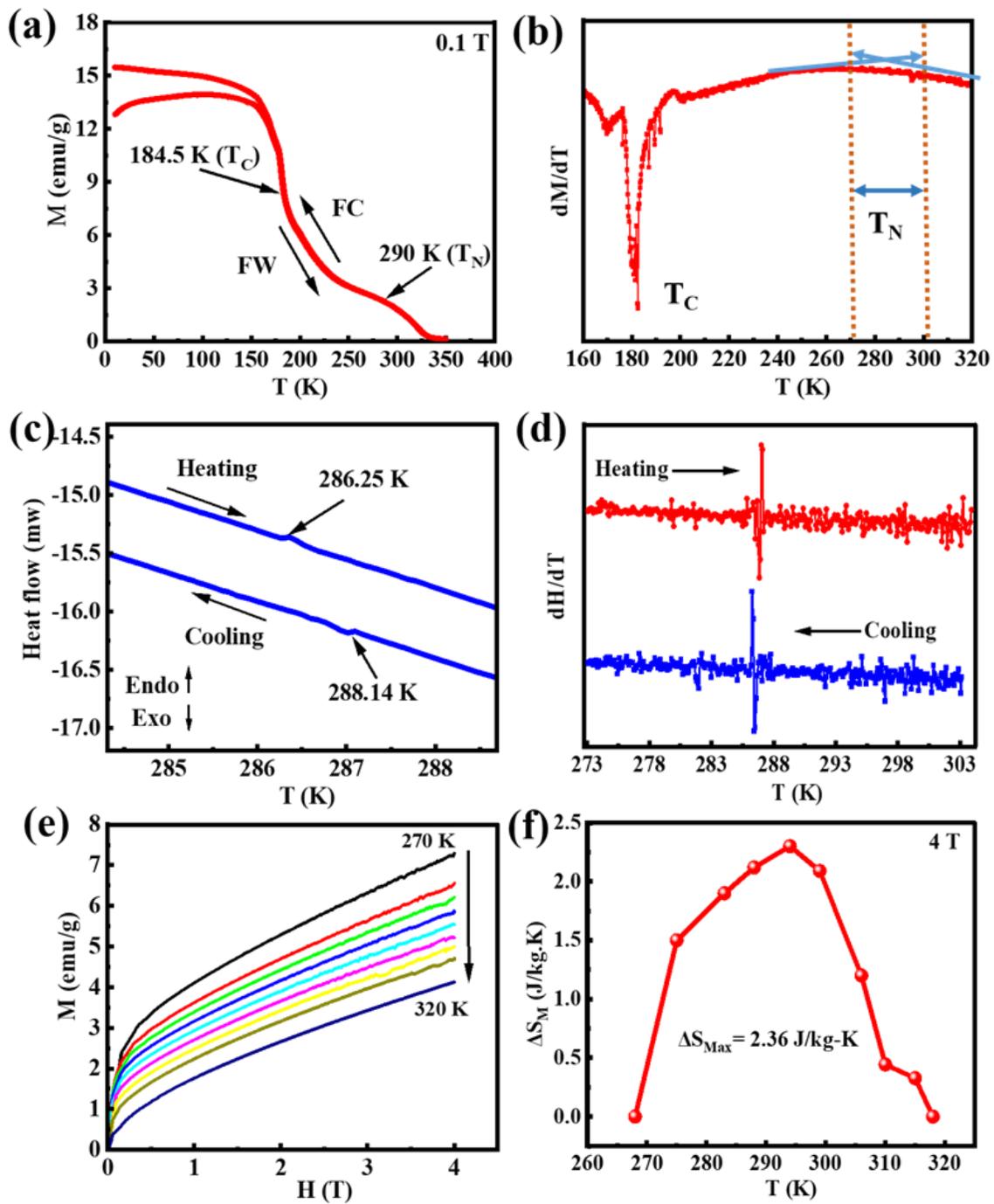

*Figure 2.* *(a) Magnetization as a function of Temperature (M-T) under the magnetic field of 0.1 T (b) First order differentiation of magnetization with respect to temperature (dM/dT) versus T (K) (c) DSC curves of $Cr_2Te_3$ in heating and cooling mode (d) First order differentiation of heat flow with respect to temperature (dH/dT) versus T (K) (e) M-H curves of $Cr_2Te_3$ across the transition*

temperature ($T_N$) (f) Calculated entropy change ($\Delta S_M$) as a function of temperature of $Cr_2Te_3$ at a magnetic field of 4 T (Tesla).

We compare these results with the existing studies involving Cr-based compositions such as $Cr_5Te_8$, $CrTe_{1-x}Se_x$, $CrI_3$, $CrB_3$, Cr-Si-Te and Cr-Ge-Te as shown in **Figure 3** [4–10,45]. In other chromium-based telluride systems, $Cr_5Te_8$ has been extensively studied. Different forms of $Cr_5Te_8$ exhibit varying values of the $\Delta S_M$. For instance, single crystal $Cr_5Te_8$ demonstrates a $\Delta S_M$ of approximately 2.42 J kg$^{-1}$ K$^{-1}$ at 322 K[4]. Comparatively, monoclinic $Cr_5Te_8$ shows a slightly lower $\Delta S_M$ of 2.38 J kg$^{-1}$ K$^{-1}$ at 225 K, and trigonal $Cr_5Te_8$ exhibits an even lower $\Delta S_M$ value of 1.93 J kg$^{-1}$ K$^{-1}$ at 240 K [5–7]. These values were all measured under a magnetic field strength of 5 Tesla, illustrating the influence of crystal structure on magnetocaloric properties. Researchers have also investigated Cr-Te-X (X = Se, Si, Ge) ternary systems, which show significant changes in magnetocaloric properties. In the Cr-Te-Se ternary system, a $\Delta S_M$ value of around 8 J kg$^{-1}$ K$^{-1}$ at 300 K was reported by varying the atomic percentages of Te and Se at 5 Tesla magnetic field[10]. In the Cr-Si-Te system, adjusting the Si and Te compositions revealed a maximum $\Delta S_M$ value of 5.05 J kg$^{-1}$ K$^{-1}$ at 30 K and 5 Tesla [45]. Beyond tellurides, magnetocaloric properties have also been explored in chromium-based halides. $CrB_3$ exhibits a $\Delta SM$ of approximately 7.2 J kg$^{-1}$ K$^{-1}$ at 33 K and 5 Tesla, while $CrI_3$ shows a $\Delta S_M$ of 5.65 J kg$^{-1}$ K$^{-1}$ under the same conditions [8,9]. This study found that $Cr_2Te_3$ has comparable values for entropy change ($\Delta S_M$) and refrigerant capacity (RC) of 2.36 J kg$^{-1}$ K$^{-1}$ and 160 J kg$^{-1}$, respectively, indicating its potential as a room temperature magnetocaloric material (MCM).

As summarized above, present study demonstrates that, in contrast to other transition metal dichalcogenides (TMDCs), the magnetocaloric properties of $Cr_2Te_3$ are comparable at room temperature under a low magnetic field. While, other TMDCs studied exhibit magnetocaloric properties that are significantly different from room temperature and require a higher magnetic field as seen in **Figure 3**.

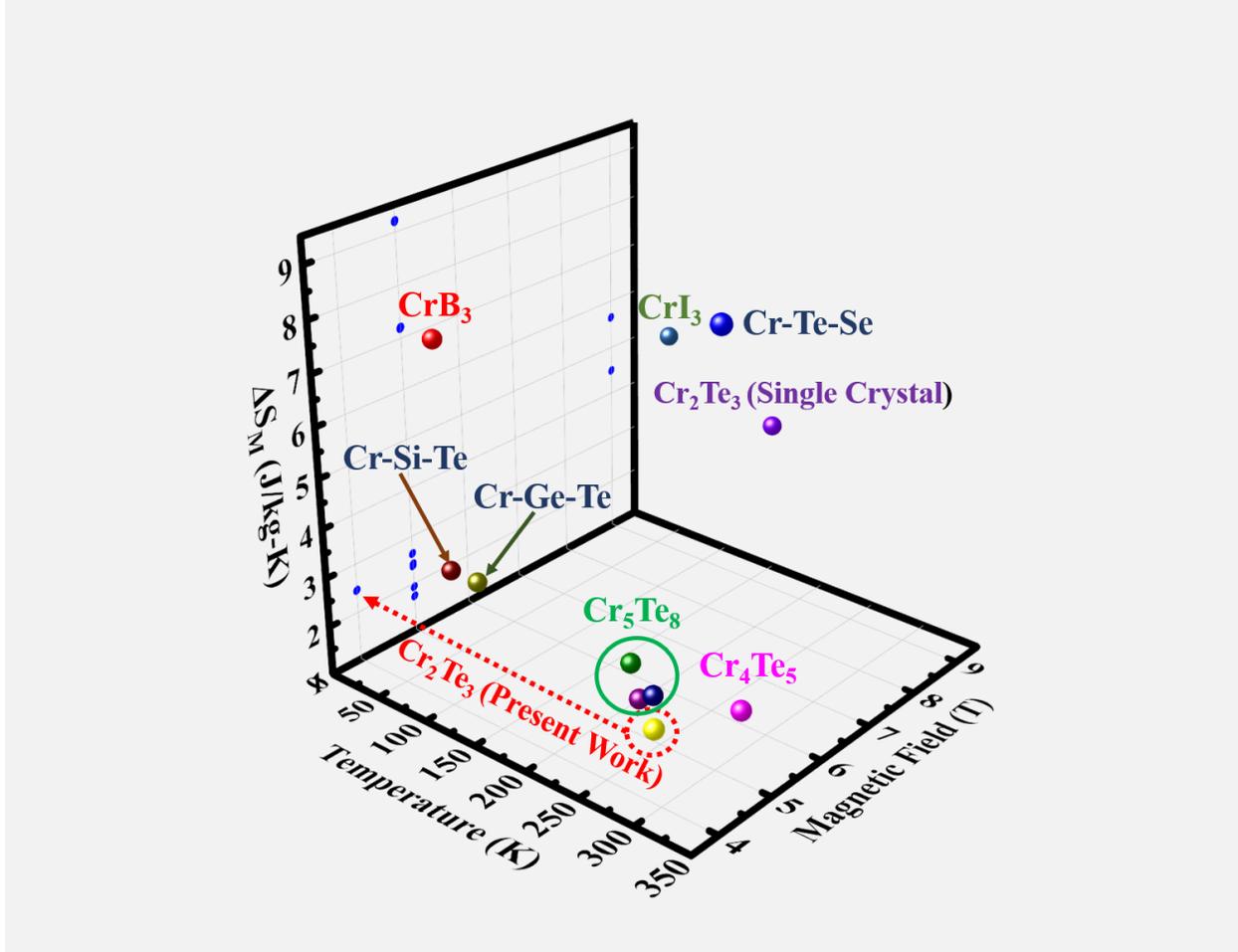

***Figure 3.*** *Magnetocaloric properties of previous studies in comparison to present study ($\Delta S_M$ ~2.36 J/kg-K) mentioned in yellow circle.*[4–10,45]

In order to validate magnetic properties, first principles calculations were performed to understand the $T_N$ (Neel Temperature) observed around room temperature for the $Cr_2Te_3$. The DFT simulated relaxed crystal structure of $Cr_2Te_3$ (top view) is shown in **Figure 4(a).** The optimized lattice parameters are $a$ = 6.90 Å, $b$ = 6.90 Å $c$ = 12.28 Å. There are 8 Cr atoms and 12 Te atoms present in the unit cell of $Cr_2Te_3$. To understand the magnetic ground state of $Cr_2Te_3$, we calculate the magnetic ground state energies of $Cr_2Te_3$ both in ferromagnetic (FM) and anti-ferromagnetic (AFM) configurations. The ground state energies in FM and AFM are -0.11800726 and -0.11729450 meV per unit cell, respectively. This indicates that $Cr_2Te_3$ exhibits FM nature at ground state.

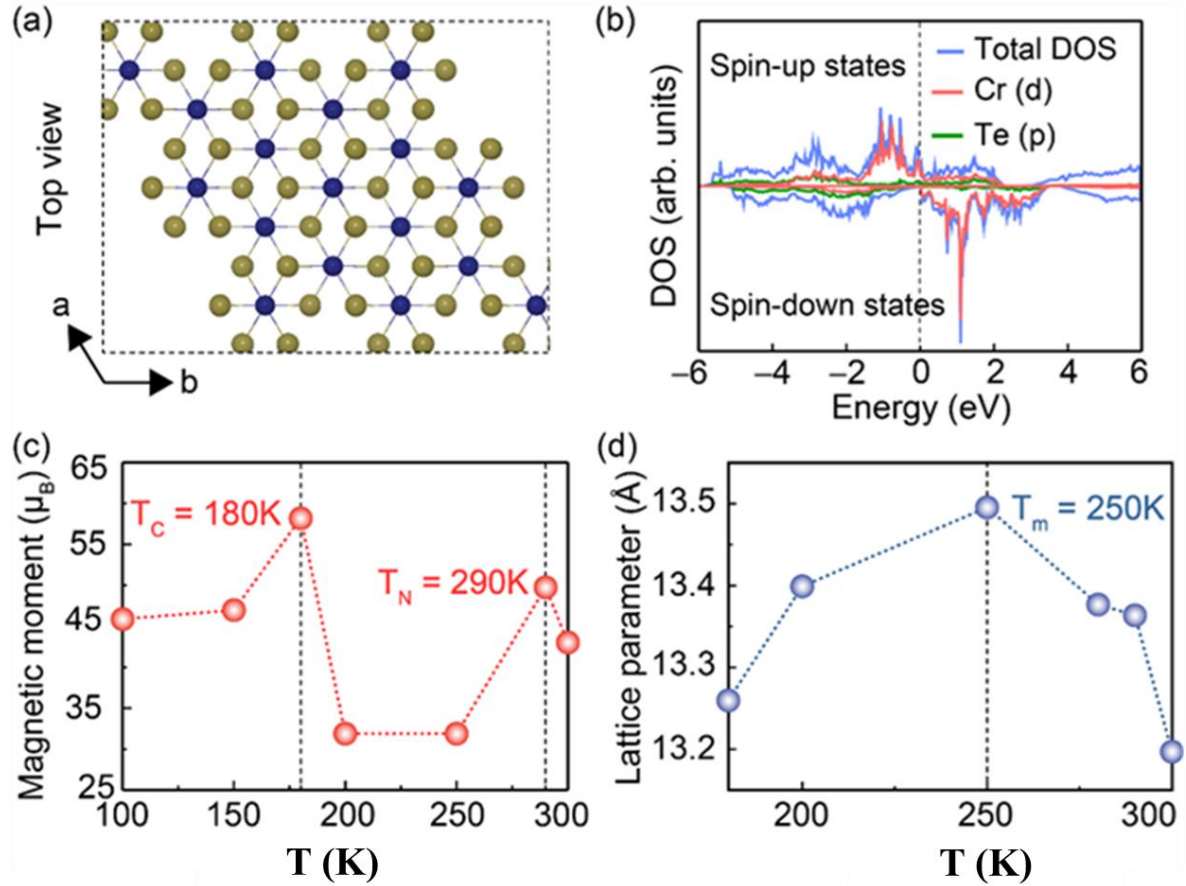

*Figure 4.* *(a) The top view of DFT simulated optimized crystal structure of $Cr_2Te_3$. (b) The spin-projected DOS for $Cr_2Te_3$. The upper and lower panel of DOS represent the spin-up and –down states, respectively. (c, d) The variation of magnetic moment and equilibrium lattice parameter of optimized $Cr_2Te_3$ at different temperatures, respectively.*

For further confirmation, we calculate the spin-projected density of states (DOS) for $Cr_2Te_3$ under FM ground state, as mentioned in **Figure 4 (b)**. Further, *d* orbitals of Cr and *p* orbitals of Te atoms mostly contribute to the total DOS. The majority (spin-up) and minority (spin-down) states are represented in the upper and lower parts of DOS, respectively. The Te atoms have almost the same number of electrons with spin-up and spin-down states, resulting in a negligible net magnetic moment. For Cr atoms, the spin-down states mostly occupy the energy levels above the Fermi level. However, the energy bands for spin-up electrons occupy the energy levels below the Fermi level. This would result in a non-zero cancellation between magnetic moment contributions from spin-up and spin-down states. Therefore, each Cr atom in the unit cell possesses a substantial magnetic moment of 3.392 $\mu_B$, which significantly enhances the system's overall magnetic moment, leading to FM ground state. To estimate the magnetic phase transition point, i.e., Curie temperature ($T_c$) of $Cr_2Te_3$, we further carried out the AIMD simulations at temperatures T = 100, 150, 180, 200, 250, 300, and 350 K. The magnetic moment variation of $Cr_2Te_3$ at different temperatures is shown in **Figure 4 (c)**. The magnitude of the magnetic moment increases with temperatures; however, after 180K, there is a sudden decrease in magnetic moment. This indicates a ferromagnetic (FM) to paramagnetic (PM) phase transition around 180 K, which is very close to the experimentally measured Curie temperature value $T_C$ = 184.5 K. Around 290 K, the magnetic moment shows a small additional increase, suggesting a possible phase transition from PM to AFM (anti-ferromagnetic) in the system. The results imply a Néel temperature ($T_N$) of 300 K, which aligns well with experimental observations of $T_N$ around 290 K. Such variation in magnetic moment values is the result of the structural change of $Cr_2Te_3$ under different temperatures. To understand the structural change, the variation of the lattice parameter of AIMD simulated crystal structure of $Cr_2Te_3$ at temperatures T = 100, 150, 180, 200, 250, 300, and 350 K is shown in **Figure 4 (d).** It can be observed that the lattice parameter increases with temperature until 250 K, where there is a sharp drop of the lattice parameter. This structural deformation results in an entropy change ($\Delta S_M$). This further suggests the structural phase transition temperature ($T_M$) at 250 K. This is very close to the experimentally measured $T_N$ value (~290 K).

Along with the theoretical validation, we have demonstrated the change in temperature under the influence of magnetic field. Visualization of the magnetic refrigeration heating was observed via a thermal imaging setup. It is well known that the magnetic refrigeration cycle consists of adiabatic followed by isothermal processes, while on the application of magnetic field two processes occur. Among that, first step is the temperature of magnetocaloric material increases under adiabatic process and subsequently the material exchanges heat at high temperature sink which is the second process (isothermal)[46]. In order to demonstrate the following increase in temperature with application of magnetic fields we incorporate a thermal imaging technique to show direct MCM property in the ferromagnetic material. The device on application of external magnetic field showed change in thermal gradient as visualized in the thermal imaging camera. The device specifications are mentioned in Materials and Methods section as shown in **Figure 5 (a)**. The device was placed on a liquid nitrogen bed without and with external magnetic field as shown in **Figure 5(b, c)**, respectively. Initially the device kept in a petri-dish temperature of low temperatures (< 0 °C) was achieved with the help of liquid nitrogen bed. **Figure 5(a)** shows the placement of device directly on the liquid nitrogen bed and temperature was raised from -35 °C to + 40 °C with the help of a hot plate. Thermal images were obtained at every 5 °C change and mentioned in **Figure 5(b, c)**. As we observe a transition near room temperature from PM to AFM at $T_N$ at 290 K (~ 17 °C), we try to observe temperature changes in the material during the transition temperature of 17 ± 5 °C. No particular trap states were observed in the material. Also, the thermal gradient between the surrounding liquid nitrogen bed and the material was at temperature gradient of ~ 2 °C consistently. **Video S1** shows heating of the material at transition temperature range from 20 °C to 30 °C (293 K to 303 K). **Fig 5(b) (i – v)** shows the gradual heating process and temperature change in $Cr_2Te_3$ in the same temperature range of 20 °C to 30 °C. The material is also placed with copper strip, another conducting element which is usually used to fabricate electro-magnetic devices. The behavior of copper remains similar with and without the application of magnetic fields.

In the next set of experiments, small magnetic fields were introduced via neodymium magnet block (1.5 Gauss) on which the sample and the coated $Cr_2Te_3$ was placed on liquid nitrogen bed. **Figure 5(c)** shows thermal gradient observed from the images which are obtained at various temperatures in presence of magnetic field. While comparing $Cr_2Te_3$ at 0 sec (20.5 °C (293.5 K)), **Figure 5(b-i) and (c-i)** shows no thermal gradient but as temperature increases the difference between thermal gradient is significant. At 300 sec (23.85 °C (296.85 K)) and 600 sec (28.6 °C (301.6 K)) time period (**Figure 5c (ii – iv)**), a large thermal gradient was observed in presence of external magnetic field. From the experimental data of temperature dependent magnetization transitions were observed at $T_N$ in the range of 288 - 296 K which led to maximum entropy change $\Delta S_M$ at around 290 K as shown in **Figure 2(a)** and **(d)**. The marked white outline in **Figure 5c** shows a few regions at lower temperatures. Along with the thermal gradient temperature difference of > 3 °C from the surroundings (magnet and liquid nitrogen bed) the images are seen to have thermal spots at lower temperatures and major regions at higher temperatures. The regions at higher temperatures are ferromagnetic cluster formation observed on application of external magnetic fields. The observation of higher temperature at a faster rate on in the presence of fields acts as a direct proof-of-concept of magnetocaloric heating cycle near room temperature.

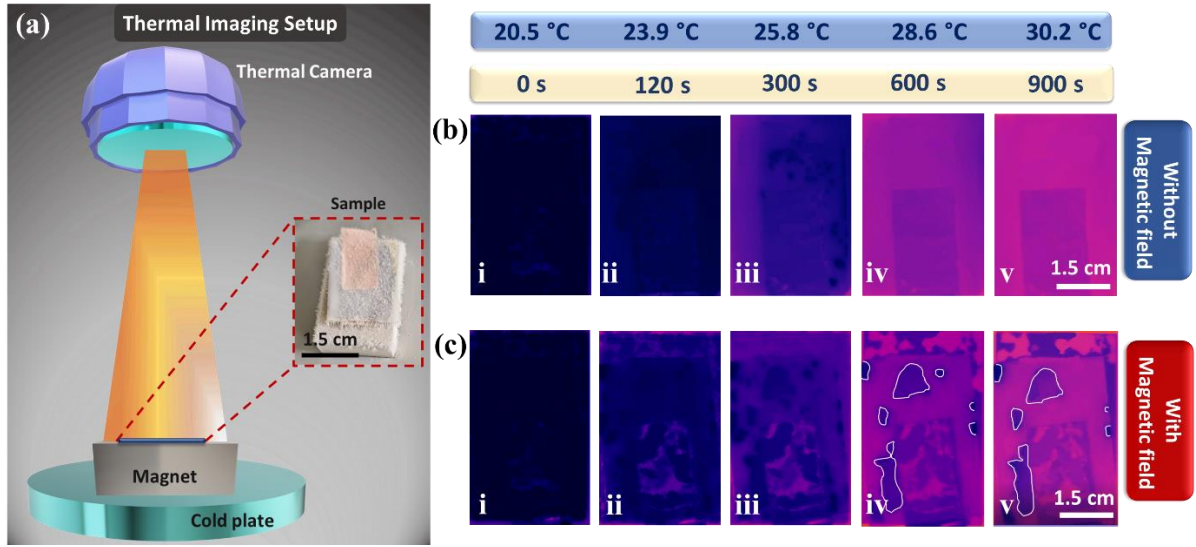

*Figure 5.* *(a) Thermal imaging camera setup, (b) Thermal images of $Cr_2Te_3$ without any external magnetic field and (c) Thermal images of $Cr_2Te_3$ with external magnetic field during gradual heating.*

On application of external magnetic field, we were able to visualize the clusters on $Cr_2Te_3$ surface through thermal imaging. The tendency to form intrinsic inhomogeneous states is observed due to ordered magnetic clusters or magnetic polarons. It has been previously observed in ferromagnetic samples such as $EuB_6$ [47] and magnetites [48]. The cluster formation was more at the critical temperature ($T_C$) and the magnetic polarons were imaged with the help of scanning tunneling microscope (STM) in these studies [47,48]. Similar cluster formation was observed in $Cr_2Te_3$ on application of external magnetic fields, as compared to images without the exposure to fields. At these points, a drastic reduction in resistance at finite magnetic fields and magnetic phase separation are observed. The suggestive magnetic clusters are strongly field dependent and effect the temperature gradient as seen in the thermal images. With higher external fields, the cluster network's magnetic moments align more and more with the direction of the field and combine with the nearby paramagnetic spins.

A potential application of this property of the material can be utilized as a heat sink on existing printed circuit boards (PCBs) [49–51]. The electric circuitry boards with in-built cooling systems have a copper coating or parallel copper plates in order to minimize localized heating of the circuit components. The copper trace thickness is provided as an impedance path for current flow. PCB management units are focusing on reducing internal heat dissipation which can manage these thermal hotspots. A coating of the ferromagnetic paste ($Cr_2Te_3$) can be proposed to replace the existing copper metallic coating due to its faster cooling rate on application/presence of external magnetic fields. The thermal resistance of the heat sink materials is often calculated using the following formula

$$R_{th} = T_1 - T_2 / P = \Delta T / P \ (°C/W) \qquad (3)$$

where ΔT stands for the temperature gradient and P for heat flow. The higher the thermal resistance the tougher the heat flow, therefore thinner materials are preferred. The MCM property of cooling on application of external magnetic field can thus be applied on a large scale for PCB and other electronic circuitry boards.

## 4. Conclusion

In summary, the magnetocaloric effect of $Cr_2Te_3$ has been investigated in detail for room temperature applications. An integrative analysis of experimental and theoretical studies was performed. The structural characterization which includes XRD, SEM, HR-TEM and XPS confirm is single phase hexagonal $Cr_2Te_3$ compound. Magnetic properties have been studied using SQUID (M-T and M-H curves). M-T curves confirm FM-PM transition in $Cr_2Te_3$ at 168 K, and the PM - AFM transition was observed around room temperature (293 K) which is an interesting aspect of the current study. DSC measurements are also conducted which in heating and cooling modes which validates the transitions observed in M-T curves of the investigated alloy. $Cr_2Te_3$ exhibits significant entropy change ($\Delta S_M$) of around 2.36 J/kg-K at a low field of 0.1 T and a refrigeration capacity (RC) of approximately 160 J/kg due to wide range of working temperature. Further, we were able to directly visualize the MCE for $Cr_2Te_3$ using thermal imaging setup. The DFT studies indicated that the lattice parameter of the relaxed structure changes with temperature, confirming the structural transformation temperature. Additionally, $T_C$ and $T_N$ was also calculated considering the FM ground state of the investigated alloys. The DOS calculations confirmed that the magnetic contribution mainly comes from Cr atoms, for the concentration studied. As the atomic positions changes with temperature which lead to change in magnetic moment, thus magnetic moment is calculated at each temperature. The magnetic moment starts decreasing at a certain point, indicating the FM to PM transition, and confirming the $T_C$ value for $Cr_2Te_3$.

AUTHOR INFORMATION

**Corresponding Author**

*Chandra Sekhar Tiwary (chandra.tiwary@metal.iitkgp.ac.in) and Abhishek K. Singh (abhishek@iisc.ac.in)

**Author Contributions**

The manuscript was written through contributions of all authors. All authors have given approval to the final version of the manuscript. ‡These authors contributed equally.

ACKNOWLEDGMENT

C.S.T acknowledges Core research grant of SERB, India, STARS projects by MHRD-India, DAE Young Scientist Research Award (DAEYSRA), and the AOARD (Asian Office of Aerospace Research and Development) grant no. FA2386-21-1-4014, and Naval research board for funding support.

ABBREVIATIONS

TMDCs, Transition metal chalcogenides; MCE, Magnetocaloric effect; MCM, Magnetocaloric materials; RC, Refrigeration capacity; DFT, Density Functional Theory; FM, Ferromagnetic; PM, Paramagnetic; AFM, Anti Ferromagnetic.

# Supplementary Information

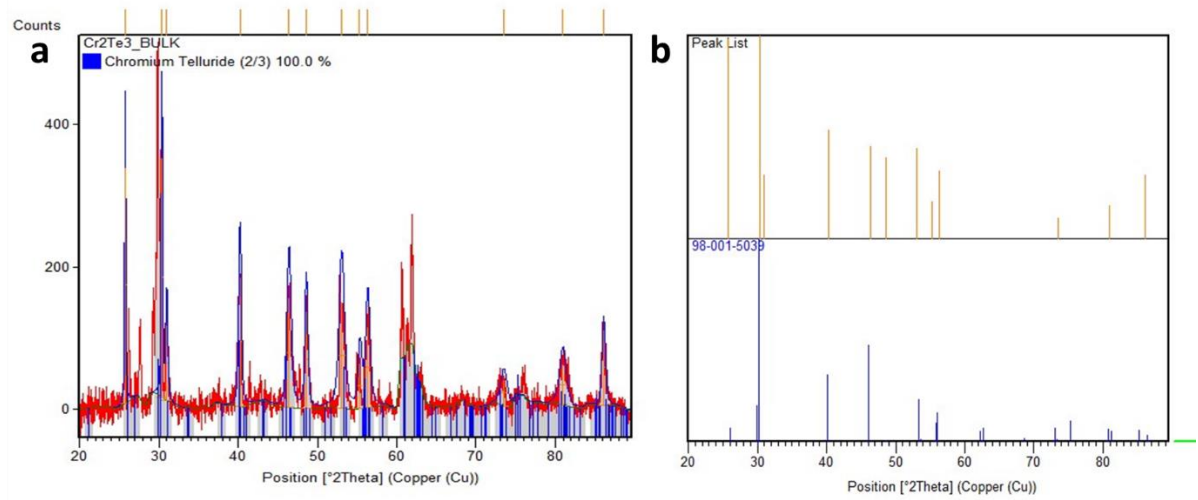

***Figure:*** *(a) Rietveld refinement of bulk $Cr_2Te_3$ sample and (b) Peaks matched with the JCPDS card number 98-001-5039.*